# Concept for the GMT High-Contrast Exoplanet Instrument GMagAO-X and the High-Contrast Phasing Testbed with MagAO-X

Laird M. Close[1a], Jared R. Males[a], Alex Hedglen[b], Antonin Bouchez[c], Olivier Guyon[a,b,d]

[a] Steward Observatory, University of Arizona, Tucson, AZ 85721, USA
[b] College of Optical Sciences, University of Arizona, Tucson, AZ 85719, USA
[c] Giant Magellan Telescope Organization (GMTO), Pasadena, CA 91107, USA
[d] Subaru Telescope, Hilo, HI, USA

## ABSTRACT

Here we review the current conceptual optical mechanical design of GMagAO-X --the extreme AO (ExAO) system for the Giant Magellan Telescope (GMT). The GMagAO-X tweeter deformable mirror (DM) design is novel in that it uses an optically distributed set of pupils that allows seven commercially available 3000 actuator BMC DMs to work "in parallel" to effectively create an ELT-scale ExAO tweeter DM – with all parts commercially available today. The GMagAO-X "parallel DM" tweeter will have 21,000 actuators to be used at ~2kHz update speeds enabling high-contrast science at ~5 mas separations in the visible and NIR of the spectrum (0.6-1.7 microns). To prove our concept for GMagAO-X several items must be lab tested: the optical/mechanical concept for the parallel DM; phasing of the GMT pupil; and solving the GMT's "isolated island effect" will all be demonstrated on an optical testbed at the University of Arizona. Here we outline the current design for this "GMT High-Contrast Testbed" that has been proposed jointly by GMTO and the University of Arizona which leverages the existing, operational, MagAO-X ExAO instrument to verify our approach to phase sensing and AO control for high-contrast GMT NGS science. We will also highlight how GMagAO-X can be mounted on the auxiliary port of the GMT and so remain gravity invariant. Since it is gravity invariant GMagAO-X can utilize a floating optical table to minimize flexure and NCP vibrations.

## 1.0 INTRODUCTION

### 1.1 Introduction to MagAO-X

To understand our approach to GMagAO-X it is important to first review our work on the MagAO-X instrument since it is the pathfinder for GMagAO-X and it is also the ExAO system for the GMT High-Contrast Testbed. MagAO-X is a unique ExAO system in that it has been targeted for primarily doing coronagraphic science in the visible part (0.5-1.0 µm) of the spectrum. This contrasts with many of today's ExAO systems that target the NIR (1-2.4 µm) for coronagraphy like GPI and SPHERE. MagAO-X leverages the excellent Las Campanas site and the slightly smaller D=6.5m size of the Magellan (Clay) telescope to allow excellent Strehls in the optical.

By use of ~1700 corrected modes (at 3.7 kHz) from an advanced woofer-tweeter design we predict Strehls of ~70% at Hα (0.6563 µm) in median seeing conditions --see Males et al. 2018 for detailed simulations of the performance of MagAO-X.

---
[1] lclose@as.arizona.edu; phone +1 520 626 5992

## 1.2 Scientific Advantages to Visible AO

Despite its demanding nature, visible AO has many scientific advantages over the NIR. After all, most astronomy is done in the visible, but almost no AO science was done with $\lambda<1\mu m$ on large 6.5-10m class telescopes until recently (Close et al. 2018). A short list of some of the advantages of AO science in the visible compared to the NIR are:

-- **Better science detectors** (CCDs): much lower dark current, lower readnoise (<1e- with EMCCDs), much better cosmetics (no bad pixels), ~40x more linear, and camera optics can be warm, simple, and compact.

-- **Much Darker skies:** the visible sky is 100-10,000x darker than the K-band sky.

-- **Strong Emission lines**: access to the primary visible recombination lines of Hydrogen (H$\alpha$ 0.6563 $\mu$m) --- most of the strongest emission lines are all in the visible, and these have the best calibrated sets of astronomical diagnostics. For example, the brightest line in the NIR is Pa$\beta$ some *20 times less strong* than H$\alpha$ (in typical Case-B recombination conditions, T~10,000K).

-- **Off the Rayleigh-Jeans tail**: Stars have much greater range of colors in the visible (wider range color magnitudes) compared the NIR which is on the Rayleigh-Jeans tail. Moreover, visible photometry combined with the IR enables extinction and spectral types to be much better estimated.

-- **Higher spatial resolution**: The 20 mas resolution regime opens up. A visible AO system at r band ($\lambda=0.62\mu m$) on a 6.5m telescope has the spatial resolution of ~20 mas (with full *uv* plane coverage unlike an interferometer) that would otherwise require a 24m ELT (like the Giant Magellan Telescope) in the K-band. So visible AO can produce ELT like NIR resolutions on today's 6.5-8m class telescopes.

## 1.3 Keys to good AO Performance in the Visible: "point design" considerations for GMagAO-X

While it is certainly clear that there are great advantages to doing AO science in the visible it is also true that there are real challenges to getting visible AO to produce even moderate Strehls on large telescopes. The biggest issue is that $r_o$ is small ~15-20 cm in the visible (since $r_o=15(\lambda/0.55)^{6/5}$ cm on 0.75" seeing site). Below we outline (in rough order of importance) the most basic requirements to have a scientifically productive visible AO system on a 6.5m sized telescope:

1. **Good 0.6" V-band Seeing Site** – Large $r_o$ (>15cm at 0.55$\mu$m) and consistency (like clear weather) is critical. In particular, low wind (long $\tau_o$ > 5ms) sites are hard to find, and so this drives loop WFS update frequencies ≥1 KHz (due in large part to the fast jet stream level winds).
→ *We could have four 2 kHz 240x240 OCAM2 PWFS to overcome this, and GMT site is 0.6" seeing.*

2. **Good DM and fast non-aliasing WFS**: need many (~21,000) actuators (with d<$r_o$/2 sampling), no illuminated "bad/stuck" actuators for GMT's D=24.5m. Currently the pyramid WFS (PWFS) is the best for NGS science (uses the full pupil's diffraction for wavefront error measurement), hence a PWFS is preferred.
→ *We have a "parallel DM" conceptual design that allows for seven 3K Boston Micromachines Corp. (BMC) DMs to be used together to effectively create the needed 21,000 actuator tweeter DM.*

3. **Minimize all non-common path (NCP) Errors:** Stiff "piggyback" design with visible science camera well coupled to the WFS –keep complex optics (like the ADC) on the common path. Keep optical design simple and as common as possible. Limit NCP errors to less than 30 nm rms. If the NCP errors are >30 nm rms employ an extra non-common path DM to fix these errors fed by a LOWFS sensor (see Males et al. 2018).
→ *On MagAO-X we have minimized the number of non-common optics to just 3 super-polished ($\lambda$/40 P-V; 0.1 nm roughness) flat mirrors, and one similar quality OAP. Hence, we expect less than ~30 nm of NCP, but have plans for an extra NCP correcting (NCPC) DM. That extra NCPC DM (another ALPAO DM-97) would be fed by a low-order WFS (LOWFS) that is directly at the coronagraphic focal plane (Miller et al. 2018). We might follow a similar approach with GMagAO-X.*

4. **Minimize the Low Wind Effect (LWE):** This is a wavefront error that is linked to the strong radiative cooling of the secondary support spiders in low wind. It can be mostly removed by adding a low emissivity coating so that the spiders track the temperature of the night air. Also, a pyramid wavefront (PWFS) sensor seems much better at sensing the LWE errors compared to SH WFS according to Milli et al. 2018.
→ *We never see LWE with MagAO currently, but if it becomes as issue with MagAO-X we expect the PWFS to sense it and the tweeter to eliminate it. We are not expecting LWE to be a problem for GMagAO-X since the GMTs spiders' geometry cannot induce it.*

5. **Minimize the Isolated Island Effect**: Unfortunately, the push towards having many corrected modes (~1700 for MagAO-X; 21,000 for GMagAO-X) forces visible AO systems to small subap sizes (13.5cm for both MagAO-

X and GMagAO-X) that approach spiders arm thicknesses (1.9 and 3.81 cm at Magellan). Hence some sub-apertures are mostly in the shadow of the spider arms and so cannot be effectively used by the WFS, allowing the DM to "run-away" in piston w.r.t. each "isolated" quadrant/section of the pupil (see Obereder et al. 2018). For GMT it is even worse in that there are large "dark" sections of the GMT pupil between the M1 segments. This is an insidious problem which favors the use of PWFSs (which could, in theory, sense the phase difference between the quadrants in these dark zones). However, it is still unproven if a PWFS can actually measure these phase differences if they are also dominated by other wavefront errors. Moreover, if the phase differences are greater than $2\pi$ the PWFS needs additional support or it will converge to the wrong (modulo $2\pi$) solution (Esposito et al. 2017) and so an additional PWFS (or other phase sensor) at another wavelength, or a real-time interferometer like a Zernike sensor (ZELDA; N'Diaye et al. 2017), or phase diversity in science focal plane; N'Diaye et al. 2018; must be used. This is not a solved problem on-sky.

→ *Extensive lab tests with realistic spider thicknesses show no sign of this effect, hence MagAO-X may not have to deal with this issue. But for GMagAO-X it will be critical to solve. We will verify that our PWFS will sense and correct these "piston" errors down to ~30 nm rms with the GMT High-Contrast Testbed feeding MagAO-X.*

4. **Lab Testing:** Lots (and lots) of "end-to-end" closed loop testing with visible science camera. Alignment must be excellent and very stiff for all non-common path optics (for all observing conditions) to minimize NCP errors.

→ *MagAO-X has already been extensively tested closed-loop in the lab. The High-Contrast Testbed will allow for us to feed a GMT pupil into MagAO-X and verify that piston or isolated island errors can be sensed and corrected (with turbulence in closed-loop) with a PWFS plus another auxiliary WFS/phase sensor operating at another wavelength than the PWFS (to eliminate the $2\pi$ phasing problem).*

5. **Modeling/Design:** Well understood error budget feeding into analytical models, must at least expect ~60 nm rms WFE on-sky. Try to measure/eliminate vibrations from the telescope and environment with advanced rejection/filter techniques (eg. linear quadratic estimation (LQG) filters).

→ *We have a full closed loop model of MagAO-X performance applied to GMagAO-X (Males et al. 2019). And we will have full Fresnel propagation of each optical surface to assess the coronagraphic performance of the system (see Lumbres et al. 2018 for the MagAO-X case).*

6 **High Quality Interaction Matrixes:** Excellent on-telescope IMATs with final/on-sky pupil. Take IMATs in partial low-order closed-loop to increase the SNR of the high order modes in the IMAT. Make the lab pupil the same as that on-sky to allow dome closed high SNR IMATs to be used later on-sky.

→ *We will be able to obtain high SNR IMATs internally with GMagAO-X and test them rigorously with our turbulence simulator in the lab or even at the telescope.*

7 **IR camera simultaneous with Visible AO camera:** this is important since you achieve a 200% efficiency boost. Allows for excellent contingency in poor seeing when only NIR science is possible.

→ *the NIR is fully corrected by GMagAO-X and will feed a future J (and maybe H band) science camera/spectrograph.*

8. **Leverage Differential Techniques for Enhanced Contrasts:** Differential techniques such as Spectra Differential Imaging (SDI) or Polarmetric Differential Imaging (PDI) are very effective in the visible, and when combined with Angular Differential Imaging (ADI) observations with Principal Component Analysis (PCA) data reduction techniques, can lead to very high contrast detections of $10^{-5}$ within 100mas (Males et al. 2016).

→ *GMagAO-X will have an excellent ADC, K-mirror, and pupil tracking loop which all together will enable long (~8 hour) pupil stabilized coronagraphic visits to targets. We will also increase contrast initially by carving a dark hole with a vAPP coronagraph (Otten et al. 2017) as done with MagAO-X (see fig 4).*

## 2.0 OPTICAL MECHANICAL DESIGN: MAGAO-X

For coronagraphy MagAO-X needs to minimize the vibrations on the optical table. We have selected a design solution where the instrument is self-contained in an enclosed floating (gravity invariant) optical table (this is the same design for GMagAO-X). The air dampening of the table will eliminate high frequency (>5Hz) vibrations from the telescope environment coupling into the instrument. However, a floating table might have low amplitude rocking at ~1Hz due to wind load etc. Hence, the rest of the MagAO-X's opto-mechanical systems (table, optical mounts, wheels etc.) are designed to be stiff enough to allow a <5 Hz solid body motion of the entire table without exciting jitter. This is critical for good coronagraphic performance. We have selected to have the actual height of the optical table to controlled, close-loop, by capacitive sensors mounted under the table. To our knowledge this is the first time this particular technical solution has been attempted at a telescope (although SPHERE has a somewhat similar system; Beuzit, J-L. private comm.).

Optically we have tried to minimize the size of the instrument, which led to a 2-tiered optical layout with some folds for compactness. We needed an excellent all reflective design so that science could be done from 0.5-1.8μm. We needed 4 pairs of super-polished OAP relays to produce 4 pupils (#1 Alpao woofer pupil (13.05mm); #2 BMC DM Tweeter pupil (18.8mm); #3 PWFS modulator/coronagraphic pupil (9.0mm); and #4 the Lyot stop pupil (9.0 mm)). We also have 5 focal planes (#1 f/11.02 telescope FP; #2 f/16.16 FP; #3 f/57 FP; #4 f/69 PWFS/Coronagraphic mask FP; #5 final f/69 post-coronagraphic science FP (6x6" FOV)). The PSF, by design, is nearly a perfect 97% SR over a very broad bandpass 0.55-0.95 microns (as seen by the PWFS) at 40-degree zenith angles thanks to our advanced ADC design. The SR is >99.5% in any one science filter.

The details of the actual design can be found below in Figs. 1, 2 and 3. The measured optical performance (closed-loop PSF and coronagraphic PSF measured in the lab) is shown in Fig. 4.

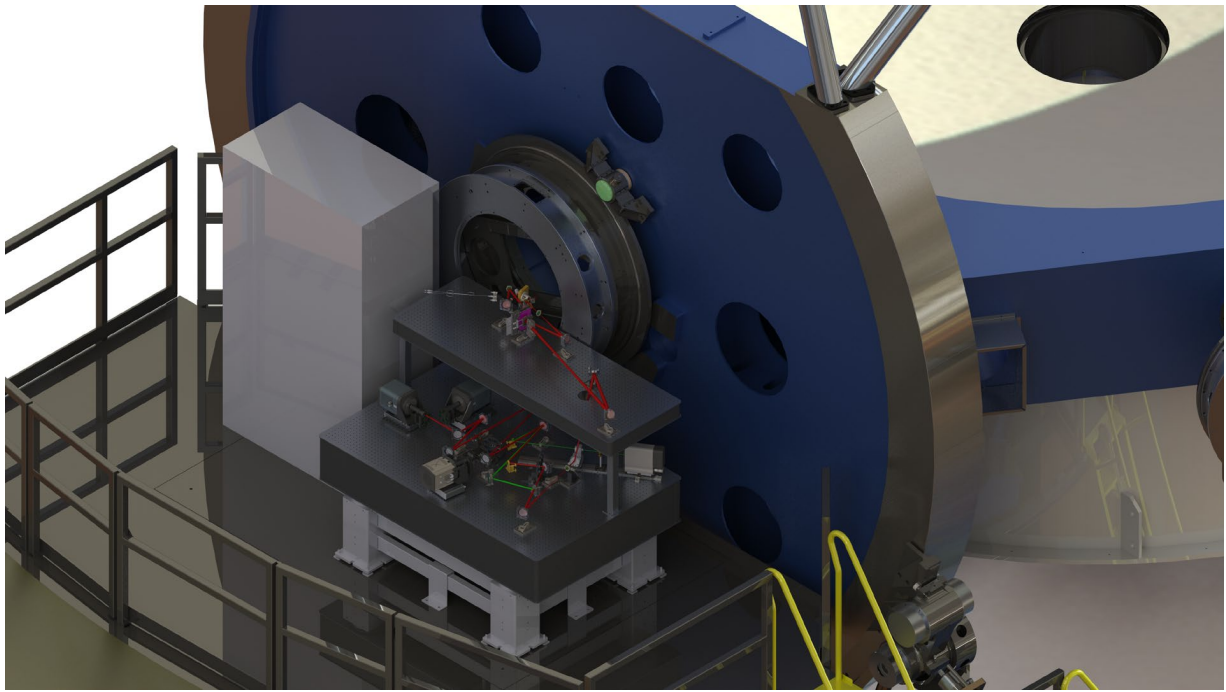

**Fig. 1:** MagAO-X as mounted at the 6.5m Magellan telescope's Nasmyth f/11 focus. The large, glycol cooled, rack to the left is for all the MagAO-X electronics. MagAO-X is gravity invariant and mounted on a floating optical table (so neither flexure nor NCP vibrations >5 Hz are issues). Note, for clarity the dust cover is removed in this rendering from MagAO-X. For more details about MagAO-X performance and science cases see Males et al. (2018).

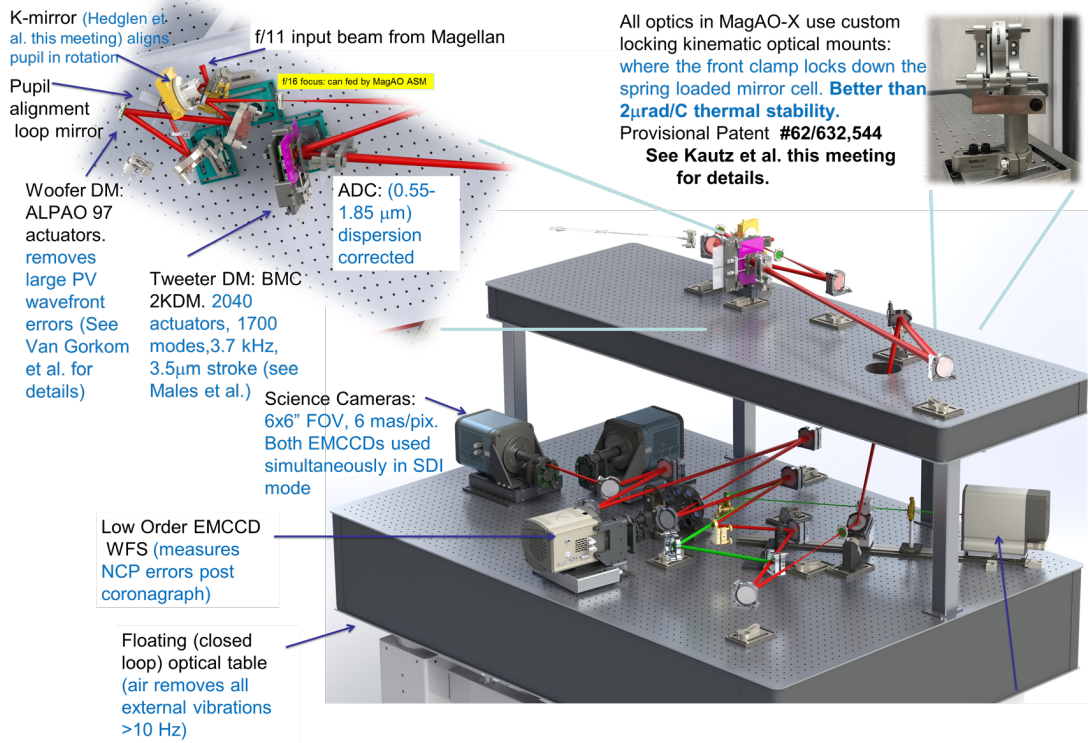

**Fig. 2:** The optical and Mechanical design for MagAO-X.

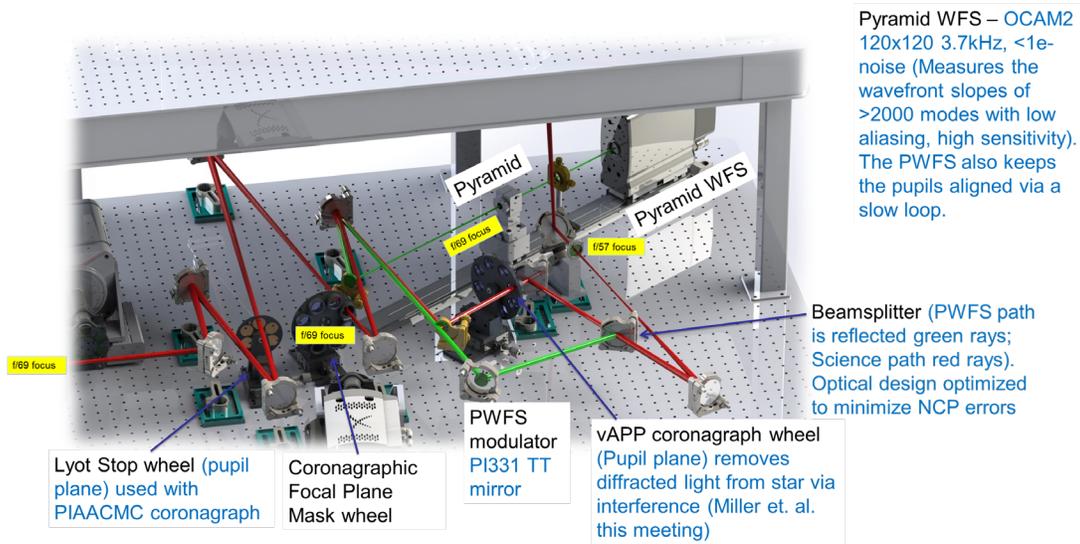

**Fig 3:** Detail of the MagAO-X vAPP Coronagraphic science (red) and PWFS (green) beam paths.

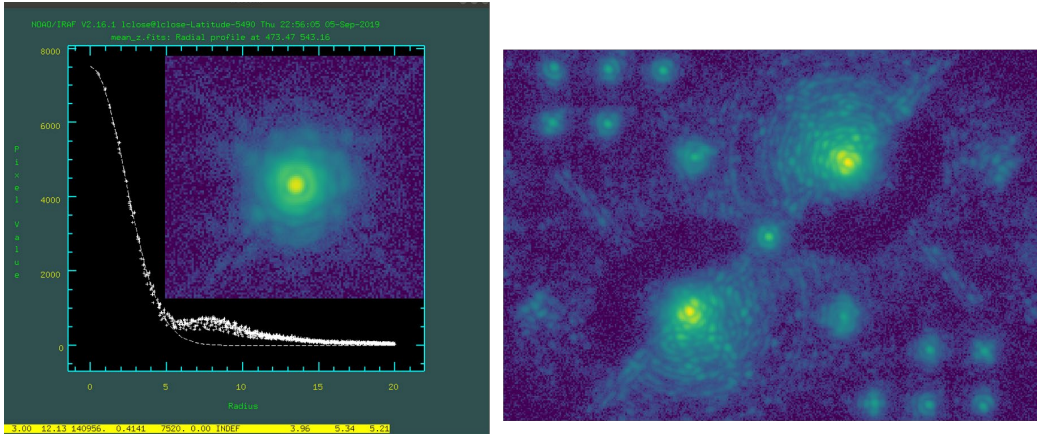

**Fig. 4:** Perfomance of MagAO-X closed-loop in the lab. Left: the z' (0.9 microns) filter PSF radial profile, with a deep log stretch insert. Right: the MagAO-X vAPP coronagraphic PSF at Halpha (0.653 microns). Note the 2 dark high-contrast dark "**C**" shaped areas that allow faint Halpha emission exoplanets to be imaged from ≥2λ/D (≥40 mas).

## 3.0 MagAO-X STATUS

During 2018/2019 in the Extreme Wavefront Control Lab (PI Jared Males) we fully assembled, tested, and documented MagAO-X. The instrument passed Magellan's pre-ship review in September and is now shipped to Chile. MagAO-X first light will be the first week of December 2019 at the 6.5m Magellan Clay telescope.

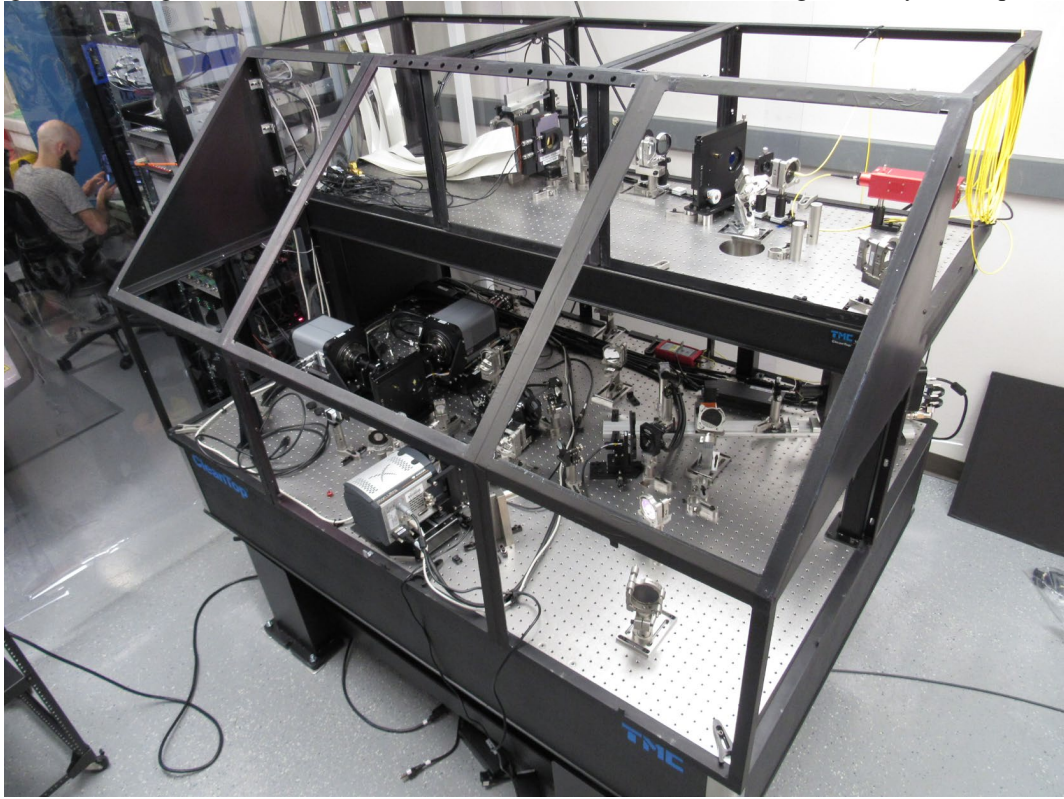

**Fig 5:** The Current (Aug. 2019) view of the MagAO-X optical table in the Clean Room Steward Observatory's High-Contrast Wavefront Sensing Lab. Kyle Van Gorkom is at the workstation that controls all of MagAO-X. Dust covers off.

# 4.0 The Giant Magellan Telescope (GMT) and GMagAO-X

The GMT is a 25m telescope composed of seven 8.4m dia. Segments (see Fig. 6). It will have first light around 2024 but without the adaptive secondaries. Here we present an ExAO instrument GMagAO-X that could achieve Strehls similar to MagAO-X but with 5 mas resolutions. GMagAO-X is an "early light" concept for a high-contrast Vis-NIR coronagraph for exoplanet and disk science. There is a description of the GMagAO-X instrument, performance, and science cases in Males et al. (2019).

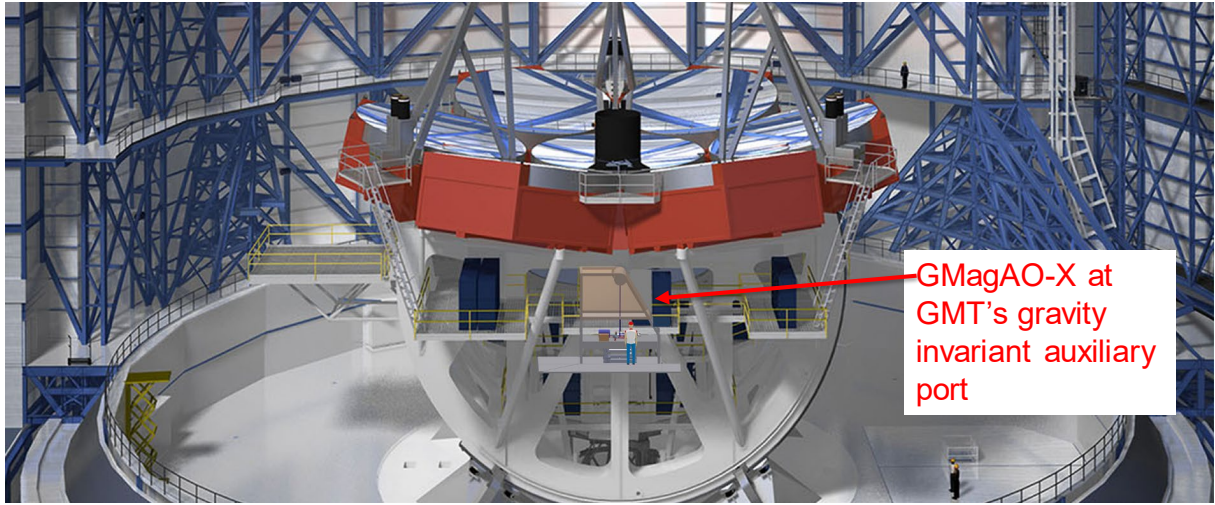

**Figure 6**: The GMT telescope with the GMagAO-X instrument concept at the Auxiliary Port (gravity invariant). At this port GMagAO-X could be a 5x10' optical table on a floating support to minimize non-common path vibrations in the coronagraph.

# 5.0 GMT High-Contrast Phasing Testbed

The scientific goals of the GMagAO-X instrument (Males et al. 2019) are to achieve very high contrasts $10^{-7}$ (ultimate goal) at very small IWA separations $3\lambda/D$ (~15 mas @ 0.65µm) in the visible and NIR (~0.6-1.5 µm). Which translates to an ExAO Strehl of 90% at 0.9µm and so requires 13.5 cm actuator pitch with a 2 KHz loop speed. A 13.5cm pitch, in turn, implies a 21,000 actuator tweeter 2 KHz DM is required. Two of the more challenging aspects of this design is: 1) we need to construct a new "Parallel DM" where 7 individual 3K BMC DMs work in parallel to produce a single 21,000 actuator tweeter DM (which is otherwise not commercially available); and 2) sensing and removing GMT M1 segment piston. To retire both of these risks we present the **GMT High Contrast Phasing Testbed** concept (see Fig. 7) that will allow us to rigorously test our sensing and control strategy for GMT NGAO and GMagAO-X in a realistic observing conditions.

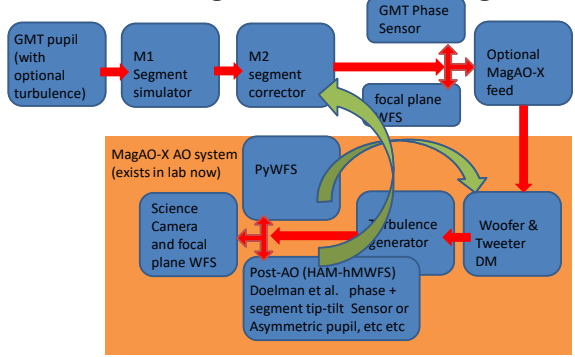

**Figure 7:** Flowchart for how the new testbed interacts with the existing MagAO-X instrument. The pyramid WFS (PyWFS) will also sense piston errors and correct them via the M2 simulator. Note that the post-AO sensors can operate at different wavelengths compared to the PyWFS hence $2\pi$ errors can be avoided

The high-contrast GMT testbed is utilizing the already existing woofer tweeter AO architecture of the MagAO-X instrument. This simplifies the hardware, software, and labor needed by a very significant margin. This leads to a cost savings for the GMT testbed on the order of 200-500%.

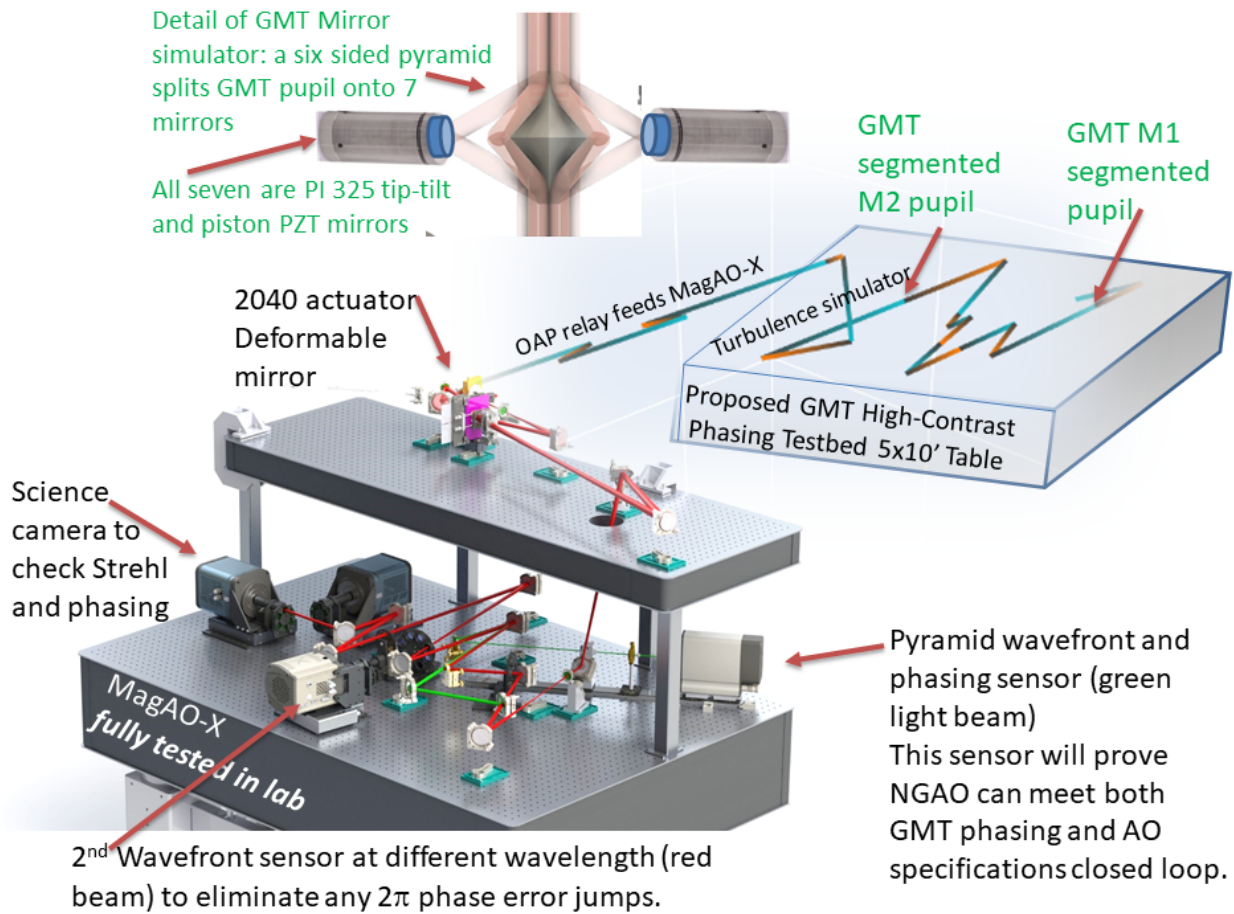

**Figure 8:** A cartoon of the GMT testbed interfacing with MagAO-X in the lab.

In Fig. 8 we show a cartoon of how the GMT testbed could be interfaced with MagAO-X. Due to the sensitive nature of tracking the phase of the wavefront as a function color we must use a reflective system. A monochromatic mode is useful for alignment purposes but ultimately we need to co-phase the test bed on the white light fringe (zero path difference between all the GMT sub-pupils). Hence, here in Fig. 9 we present an initial all reflective (OAP based) design for the GMT testbed.

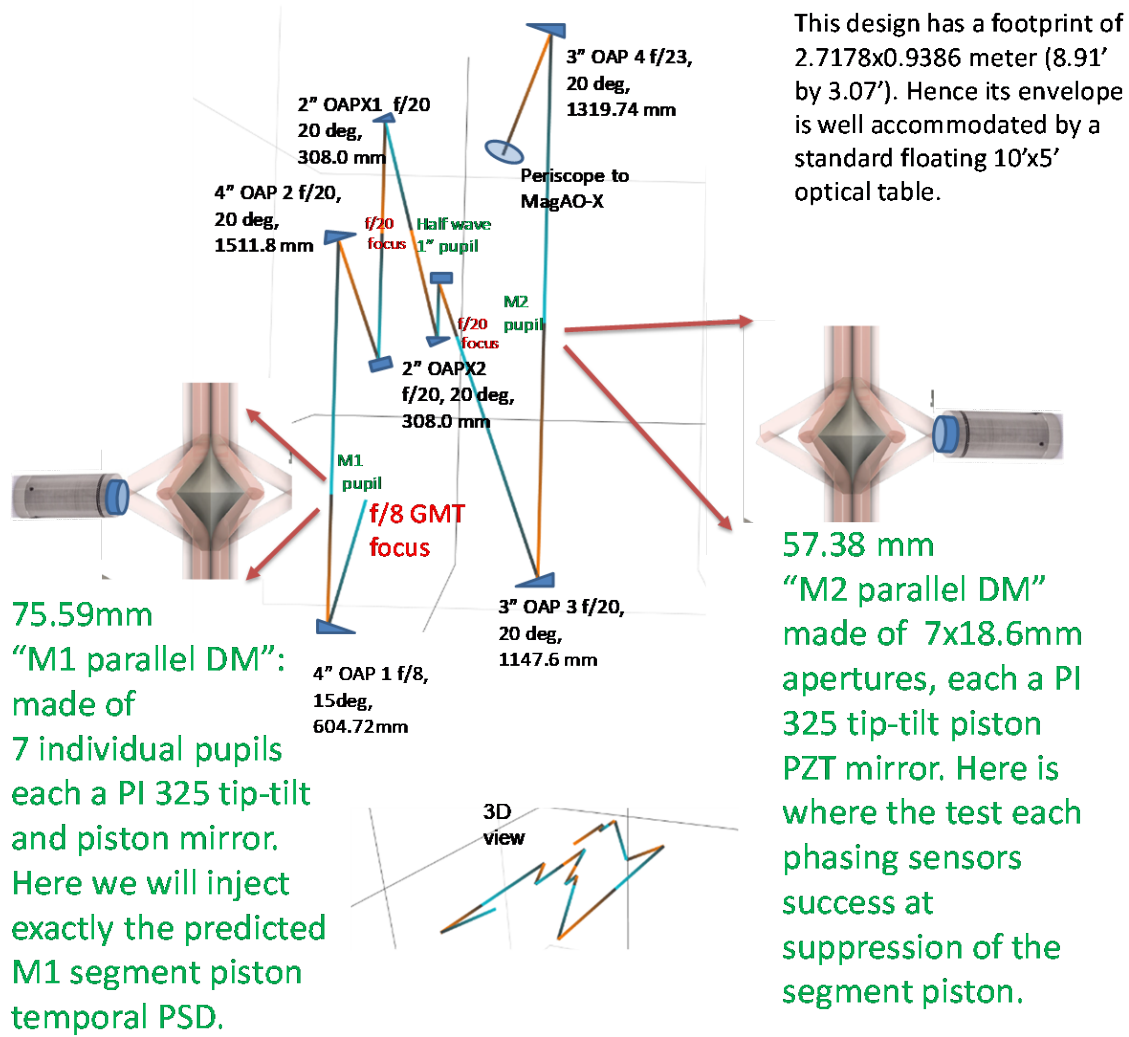

**Figure 9:** An all reflective design for the testbed concept using OAPs to relay the pupils and focal planes.

## 6.0 A Parallel DM for GMagAO-X

Ultimately, we would like this testbed to prove a concept for extending extreme AO to ELT sized apertures. This is a problem for ELT, TMT and GMT in that the >20,000 actuators required for an ExAO ELT class instrument sis simply not commercially available today. So here we present the concept of a "Parallel DM", where even though no one commercial DM can achieve the fitting error required, we can use, say 7 commercial 3,000 element DMs (like the BMC 3K device) in parallel and achieve a functional 21,000 actuator "parallel DM" which would function as a single 21,000 actuator DM. This is the baseline for the GMT's ExAO spectrograph/coronagraph GMagAO-X (see Males et al. these proceedings). There are some slight optomechanical considerations to achieve a highly co-phased (coherent) output pupil identical to the input pupil that need to be tested for the Parallel DM.

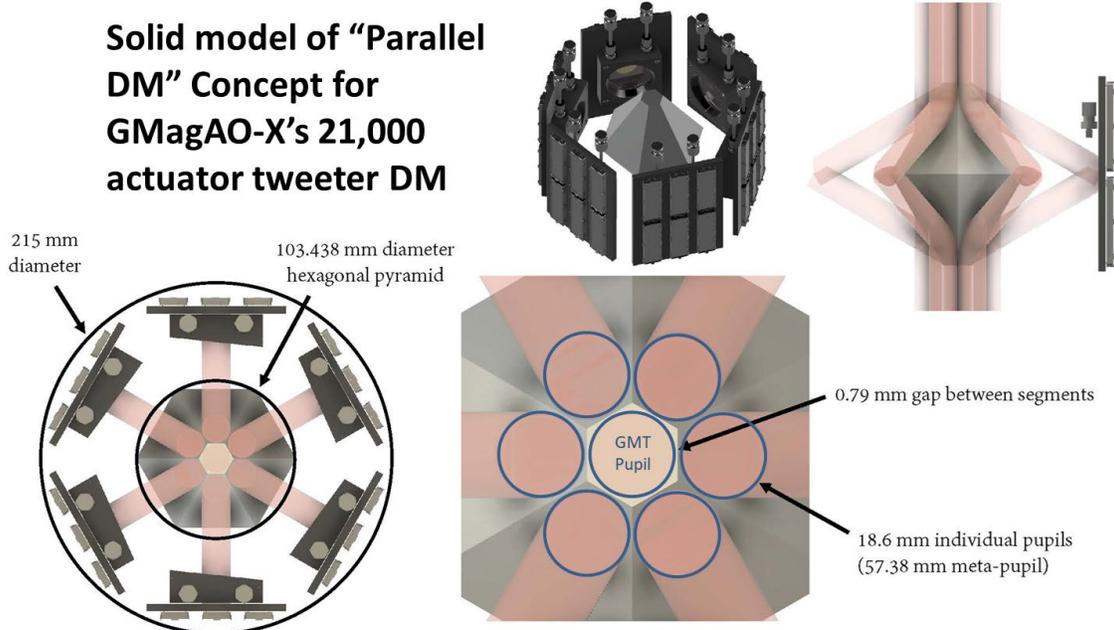

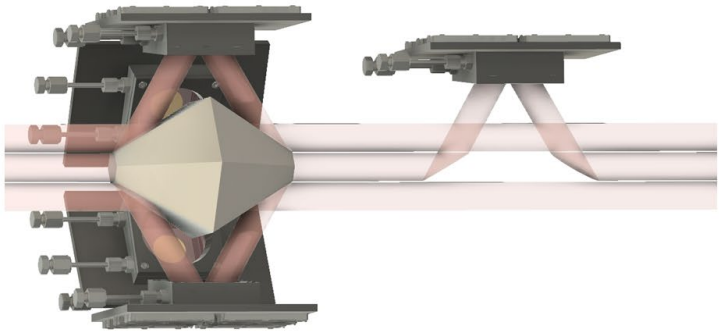

**Figure 10:** A solid model view of how the segmented nature of the GMT pupil allows for seven 3K DMs (six split into a ring and the center 7th sub-pupil passes straight through the 6-sided pyramid optic) to act effectively as a 21,000 actuator, coherent, tweeter DM.

The parallel DM concept will be completely tested on the GMT testbed. This will test for the stability and coherence of this approach. Obviously, a key sensing issue to keep the "isolated island" effect to a minimum and keep all 7 pupils co-phased. This will rely on the standard dispersed fringe sensor (DFS) of the GMT (Kopon et al. 2018; and these proceedings), which can capture as much as 30 microns of segment piston error. However, at the ~20-30 nm rms error needed for true ExAO performance we need a more sensitive piston sensor. We propose that the Pyramid sensor of MagAO-X should be able to measure very small phase errors once the errors have been reduced to <lambda/2 by the DFS.

This piston sensing and correction will be a key performance goal for the GMT phasing testbed to demonstrate. We can drive into the M1 simulator to the expected piston and tip-tilt vibrational power spectra (See Bouchez et al. these proceedings). While these errors are being driven into the PSD of the M1 PI S-325 actuators, the PyWFS will feed back the correct tip-tilt and piston settings on the M2 simulator's S-325s. In this manner, realistic errors that will arise on M1's segments will be measured and corrected with M2's segments.

## 7.0 SUMMARY

**MagAO-X** has passed its pre-ship review and demonstrated closed-loop "mock" coronagraphic science operations on the 97 actuator Woofer and 2048 act. BMC DM. It is currently shipped to Chile and on schedule for first light at the Magellan 6.5m Clay telescope in Dec. 2019. See Males et al. 2018, 2019, Close et al. 2018.

The **GMT High-Contrast Phasing Testbed** is at the conceptual phase and will be directly feeding MagAO-X (in between telescope runs) at Steward Observatory, University of Arizona and will be funded under contract by GMTO. It will work with MagAO-X to prove how AO and segment piston sensing can work at high-contrast with a PyWFS and the GMT pupil as well verifying the parallel DM optomechanical design.

**GMagAO-X** is a future "early 2nd gen." high-contrast Vis-NIR exoplanet imager and spectrograph for the GMT. The instrument is at the conceptual level (see Males et al. 2019 for details about the science cases and performance goals) and will use results from the high-contrast testbed to move on to the PDR level design.

MagAO-X could not have been possible without support from the NSF's Major Research Infrastructure (MRI) research grant # *1625441 Development of a Visible Wavelength Extreme Adaptive Optics Coronagraphic Imager for the 6.5-meter Magellan Telescope* (PI: Jared Males). L.M.C.'s research was also supported by the NSF AAG program #1615408 (PI Laird Close). Laird Close and Alex Hedglen are supported by the NSF's AAG program #1615408 and NASA's XRP program 80NSSC18K0441 (PI Laird Close).